\documentclass{appolb}
\usepackage{epsfig}
\begin{document}
% \eqsec  % uncomment this line to get equations numbered by (sec.num)
\title{NONEXTENSIVE INFORMATION ENTROPY FOR STOCHASTIC NETWORKS}
\author{G.Wilk\thanks{e-mail: wilk@fuw.edu.pl}
\address{The Andrzej So\l tan Institute of Nuclear Studies, 
         Ho\.za 69; 00-689 Warsaw, Poland}
\and  Z. W\l odarczyk\thanks{wlod@pu.kielce.pl}
\address{Institute of Physics, \'Swi\c{e}tokrzyska Academy,
         \'Swi\c{e}tokrzyska 15; 25-406 Kielce, Poland\\ 
         and\\
         The Kielce School of Science (WSU), Weso\l a 52; 
         25-353 Kielce, Poland
}
}
\maketitle

\begin{abstract}
Nature is full of random networks of complex topology describing
such apparently disparate systems as biological, economical or
informatical ones. Their most characteristic feature is the apparent
scale-free character of interconnections between nodes. Using an
information theory approach, we show that maximalization of 
information entropy leads to a wide spectrum of possible types of
distributions including, in the case of nonextensive information
entropy, the power-like scale-free distributions characteristic of
complex systems.\\    

\noindent
PACS numbers: 89.75.-k 89.70.+c 24.60.-k \\
{\it Keywords:} Complex networks, Information theory, Nonextensive
statistics\\  
\end{abstract}

Random networks have recently found applications in the description
of complex systems in different, apparently very disparate branches
of modern science such as, for example, molecular biology, sociology,
economy and computer science \cite{GA,Service}. For example, living
organisms form huge genetic networks the nodes of which are proteins
and links represent the corresponding chemical interactions
\cite{WBI}. A similarly big network is formed by the nervous system
the nodes of which are connected by axions \cite{KL}. Comparable
complexity show networks existing in the sociological systems in
which nodes are countries, organizations or single persons whereas
links characterize their mutual interactions \cite{WF}, in the world
of finances and computer networks (with World Wide Web being the most
known example where nodes are HTML documents connected via
hiper-links URL \cite{AJB}). For most recent reviews of random
networks see \cite{AB,DM}.\\ 

Analysis of different random networks clearly indicate that the
probability $P(k)$ of joining a given node with other nodes is 
described by the power law $P(k) \propto k^{-\gamma}$ \cite{BA}.
For example, the most convincing analyses of computer networks
with over $800$ milion nodes \cite{LG,AJB,BAJ,JK} lead to a 
power-like distribution of $P(k)$ with exponent equal $\gamma \sim
2.1\div 2.45$. This contradicts the existing models of random
networks \cite{ER,B} predicting instead exponential distributions:
$P(k) \propto \exp (-k)$. The most popular model (ER) dealing with
a fixed number of nodes $N$ was proposed in \cite{ER}, where the
Poisson distribution was advocated to be used for probability that a
given node has $k$ links (with the mean number of connections being
$\lambda_0$) \footnote{Notice that in the limit of large $k$
distribution (\ref{eq:ER}) can be approximated by $P(k) = (2\pi
k)^{-1/2}\left(\lambda_0/k\right)^k \exp\left( k - \lambda_0\right)$,
whereas for large values of $\lambda_0$ it becomes a gaussian
distribution, $P(k) = \left(2\pi \lambda_0\right)^{-1/2}\exp\left[
-\left( k - \lambda_0\right)^2/2\lambda_0\right]$.},
\begin{equation}
P(k)\, =\, \frac{\lambda_0^k\, e^{- \lambda_0}}{k!} . \label{eq:ER}
\end{equation}
However, in order to get the observed power-like form of $P(k)$ one
has to allow for growing $N$ and replace the democratic law of
attachement a new link used in deriving (\ref{eq:ER}) by a
preferential one. This means that distribution $P(k)$ is determined 
by the dynamics of the growth of network \cite{AB,BAJ}. Starting from
a small number $m_0$ of nodes, adding in each time step a new node
with $m \leq m_0$ possible connections and assuming that this new
node 
joins the already existing nodes with 
equal, $k_i$-independent, probability $\Pi (k_i)=\frac{1}{m_0 +
t - 1}$, the evolution (growth) of network is described by the
following equation \cite{BAJ}: 
\begin{equation}
\frac{\partial k_i}{\partial t}\, =\, m\, \Pi (k_i)\, =\,
\frac{m}{m_0 + t - 1} \label{eq:Part}
\end{equation}
leading for long times $t$ to exponential stationary distribution:
\begin{equation}
P(k) \, =\, \frac{1}{m} \exp \left( - \frac{k}{m}\right) .
\label{eq:Pois} 
\end{equation}
On the other hand, assuming that probability $\Pi (k_i)$ is
selective, for example that $\Pi (k_i) = 
\frac{k_i}{\sum^{m_0+t-1}_{j=1}k_j}= \frac{k_i}{2mt}$,
one gets instead asymptotically a simple power law for $P(k)$:
\begin{equation}
P(k)\, =\, \frac{2m^2t}{m_0 + t}\, k^{-3}\, \propto\, k^{-3} .
\label{eq:-3} 
\end{equation}
Apparently such distribution with universal exponent $\gamma = 3$
shows up in different situations (and under different names). As 
Pareto distribution \cite{P} it describes the growth of the wealth of
persons living in stable economical systems, as Zipf's law \cite{Z} it
is applied in linguistic and  it also describes the distribution of
the citations of the scientific works \cite{R,TA}.\\

In the limit of large $t$, i.e., when stationary state already
develops, this problem can be also studied from the information
theory point of view. In it one asks the following question
\cite{ME}: what is the informational content of data represented by
distributions $P(k)$? In other words, what is the minimal number of
parameters needed to reproduce a given shape of $P(k)$? The question
asked in such approach is: suppose that we know only that a network
we are interested in, which we would like to describe by $P(k)$,
leads to some  mean value of $k$, i.e., we know that: 
\begin{equation}
\sum^{\infty}_{k=1}\, P(k)\, =\, 1\qquad {\rm and}\qquad \langle
k\rangle\, =\, \sum^{\infty}_{k=1}\, k\cdot P(k)\, =\, \lambda_0\, =\,
const , \label{eq:cond}
\end{equation}
what would be then {\it the most probable and least biased} form of
$P(k)$ in such a case (i.e., describing the existing data and given
entirely in terms of $\lambda_0$)? To answer this question one
maximalizes the corresponding information entropy associated with
probability distribution $P(k)$ under constraints (\ref{eq:cond}).
The usual form of such entropy is Shannon entropy \cite{Shan}, 
\begin{equation}
S\, =\, - \sum^{\infty}_{k=1}\, P(k)\, \ln P(k) . \label{eq:Shan}
\end{equation}
The conditions (\ref{eq:cond}) representing our {\it a priori}
knowledge of the problem lead to the exponential probability
distribution 
\begin{equation}
P(k)\, =\, \frac{1}{\lambda_0}\cdot \exp\left( -
\frac{k}{\lambda_0}\right) , \label{eq:exp}
\end{equation}
closely resembling eq.(\ref{eq:Pois}) \footnote{It is interesing to
mention that additional knowledge that all entities represented by
$k$ are {\it indistinguishable}, which results in the necessity of
introducing in the correspodning sumations the additional weight
factor $1/k!$, changes the above distribution to Poisson distribution
of the ER model mentioned above, cf., eq.(\ref{eq:ER}).}.\\

There are, however, many systems with properties preventing the use
of Shannon type of information entropy and calling for its
generalization. For example, they posses some intrinsic fluctuations
resulting in the whole spectrum of parametr $\lambda$, $P(\lambda)$,
with $\lambda_0 = \langle \lambda \rangle$ being only its mean value
\cite{WW} or they develop some correlations introducing memory
effects, cf. \cite{T} (in statistical physics such situation leads to
the necessity of departing from the use of the usual Boltzmann-Gibbs
statistics in favour of some sort of generalized one \cite{T}). Out
of many possible generalizations we shall use in this note the
nonextensive Tsallis entropy defined as \cite{T}:    
\begin{equation}
S_q\, =\, - \frac{1}{1-q}\cdot \left\{1\, -\, \sum^{\infty}_{k=1}\,
\left[P(k)\right]^q\right\}. \label{eq:Sq}
\end{equation}
It can be regarded as a minimal (i.e., one parameter) extension of
Shannon entropy (\ref{eq:Shan}), to which it reduces when
$q\rightarrow 1$. Parameter $q$ describes therefore summarily all
effects preventing the use of Shannon entropy mentioned above.\\

Using $S_q$ as a measure of information about our system, i.e.,
maximalizing $S_q$ with constraints (equivalent to (\ref{eq:cond})
above):
\begin{equation}
\sum^{\infty}_{k=1}\, P(k)\, =\, 1\qquad {\rm and}\qquad \langle
k\rangle_q\, =\, \frac{\sum^{\infty}_{k=1}\, k\cdot [P(k)]^q}
                      {\sum^{\infty}_{k=1}\, [P(k)]^q} \, =\, \lambda_0
\, =\, const , \label{eq:condq}
\end{equation}
one obtains as result a  power-like distribution of the form:
\begin{equation}
P(k)\, =\, P_q(k)\, =\, C\cdot \left[1\, -\, (1-q)\cdot
\frac{k}{\lambda_0}\right]^{\frac{q}{1-q}} , \label{eq:Pq}
\end{equation}
where $C = 1/\sum^{\infty}_{k=1}[1-(1-q)k/\lambda_0]^{q/(1-q)} =
1/\lambda_0$ is normalization\footnote{It should be stressed that
maximalization of entropy provides us in this case only with the
shape of distribution $P(k)$, eq. (\ref{eq:Pq}), and gives no
information on  the particular values of parameters $\lambda_0$ and
$q$. Only knowledge of moments $\langle k\rangle$ and $Var(k)$ of
$P_q(k)$, as given by eq. (\ref{eq:moments}), allows for determining
these two parameters.}. In this case 
\begin{equation}
\langle k\rangle = \frac{\lambda_0}{(2-q)}\quad {\rm and}\quad 
Var (k) = \frac{\lambda_0^2}{(3-2q)(2-q)^2} . \label{eq:moments}
\end{equation}
Notice that for $q\rightarrow  1$ this distribution becomes
exponential, as in eq.(\ref{eq:exp}). On the other hand, for large
values of $k$, $k >> \lambda_0/(q-1)$, it becomes a power-like distribution of the form
\begin{equation}
P_q(k)\, \propto\, k^{-\gamma}\qquad {\rm with}\qquad \gamma
=\frac{q}{q-1} , \label{eq:power}  
\end{equation}
i.e., our distribution becomes in this limit a scale-free one.
It is easy to check that if we demand that $\langle k\rangle <
\infty$ then $q<2$. It is interesting to note at this point that
$\gamma = 3$ in eq.(\ref{eq:-3}) corresponds precisely to $q=3/2$ at
which variation $Var (k)$ diverges.

\vspace{-7mm}
\begin{figure}[ht]
\begin{center}
        \epsfig{file=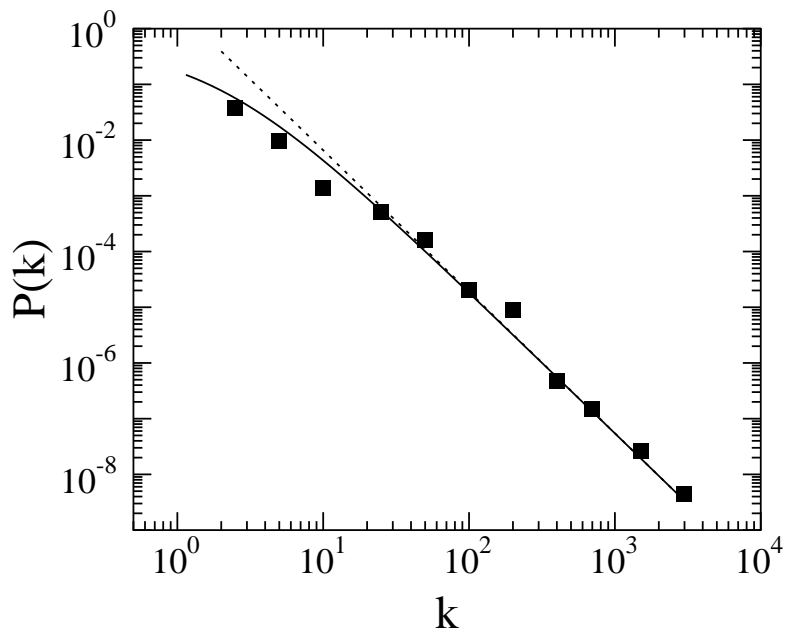, width=120mm}
\end{center}
  \caption{The probability distribution of connections in the WWW
           network after \protect\cite{JK} (full squares). The full
           line shows results of our fit by using eq. (\ref{eq:Pq}) 
           with $q=1.65$ and $\lambda_0 = 1.91$. It reproduces the
           observed mean $\langle k\rangle = \lambda_0 /(2-q) = 5.46$
           and lead to the asymptotic power-like distribution
           $\propto k^{-\gamma}$ with $\gamma =q/(q-1) = 2.54$
           (showed as dotted line). }
  \label{Figure1}
\end{figure}

In Fig. 1 we show (as example) distribution of the number of
connections in the WWW network \cite{JK} with $325729$ nodes and the
mean values of connections equal $\langle k\rangle = 5.46$ fitted by
$P_q(k)$ as given by eq.(\ref{eq:Pq}) with parameters $\lambda_0 =
1.91$ and $q=1.65$ \footnote{Notice that although in fitting
procedure both parameters were varied independently, they are
connected via distribution moments $\langle k\rangle$ and $Var(k)$,
cf. (\ref{eq:moments}).}. Notice that eq.(\ref{eq:Pq}) describes {\it
the whole range} of $k$ whereas the purely power-like distribution
$\propto k^{-\gamma}$ with $\gamma =q/(q-1) = 2.54$ occurs only for
large values of $k$. In the spirit of information theory this result
can therefore be interpreted in the following way: $(a)$ the system
forming network described in Fig. 1 posseses some features (mentioned
above) preventing the use of Shannon information entropy and $(b)$
data represented by $P(k)$ can be quite adequately described in terms
of only two parameters: the Tsallis entropy parameter $q$ and the
mean number of links $\langle k\rangle_q = \lambda_0$, i.e., their
informational capacity is rather limited. \\ 

The question now is: what is the physical meaning of the $q$
parameter in the context of stochastic networks? There is a long list
of possibilities in what concerns of the origin of $q\neq 1$ to be
found in the literature dealing with nonextensivity \cite{T,WW}. Out
of these we shall only mention two: fluctuations and correlations. In
\cite{WW} it was demonstrated that $q$ reflects fluctuations of the 
parameter $\lambda_0 $ in exponential distribution (\ref{eq:ER})
above. In fact, it turns out that $(q-1)/q = \pm
Var(1/\lambda)/\langle 1/\lambda\rangle^2$. As is known from other
branches of physics where Tsallis statistics is applied \cite{T}, the
appearance of $q$ can also be caused by some correlations existing in
the system under consideration. Such correlations (resulting, for
example, from preferential attachements and "rich-get-richer"
phenomenon \cite{BAJ}) seem to play a decisive role in the
description of stochastic networks. Therefore when choosing vertices
with connectivity $k$, to which a new vertex is going to be
connected, we shall assume that it will do so with probability that
depends on the connectivity $k$. To illustrate this point let us
introduce in the evolution equation
\begin{equation}
\frac{dP(k)}{dk}\, =\, - \frac{1}{\lambda(k)} P(k), \label{eq:dPk}
\end{equation}
parameter $\lambda =\lambda(k)$ given by a simple linear function of
$k$:  
\begin{equation}
\lambda(k) = \frac{\left[ \lambda_0 + (q -1)k\right]}{q} . \label{eq:L}
\end{equation}
It is easy to see that in this case one gets immediately $P_q(k)$ in
the form of eq. (\ref{eq:Pq}). Notice that for $q\rightarrow 1$
(i.e., for $\lambda \rightarrow \lambda_0$) one recovers the
exponential distribution (\ref{eq:exp}) \footnote{Eq. (\ref{eq:dPk})
can be derived by dividing master equation, $\partial P(k)/\partial t
= - c\, P(k)$, by the "growth of network" $\partial k/\partial t$.
One gets then evolution equation, $\partial P(k)/\partial
k = - c\, P(k) \partial t/\partial k$, which for the linear
dependence of the grow of network assumed here, $\partial k/\partial
t = a + bk$, leads to eq. (\ref{eq:dPk}) with $\lambda (k)
=(a+bk)/c$. Our example considered here corresponds to the choice:
$c=1/\lambda_0$, $a=1/q$ and $b=(1-1/q)/\lambda_0$. Notice that $1/q$
plays role of weight with which we select the constant and linear
terms in the equation describing the growth of network. Notice also
that this kind of the growth of network, i.e., its dependence on $k$
(cf. \cite{BAJ}) corresponds to selective probability $\Pi (k)$,
which leads to power-like distribution (\ref{eq:-3}).}. \\  

It must be stressed, however, that the information theory approach
leads in a natural way (via maximalization of the respective
information entropy) only to equilibrium (or stationary)
distributions $P_q(k)$, whereas in models describing evolving complex
networks \cite{BAJ,AB} the functional form of $P(k)$ is determined by
the growth equation $\partial k/\partial t$. Introducing more
complicated network evolution than the one presented above when
deriving eq.(\ref{eq:-3}), for example allowing for the occurence of
local events in the form of internal edges and rewirings, one gets
(see eq. (111) of \cite{AB})  
\begin{equation}
P(k) \sim \left[k + \kappa(p,r,m)\right]^{-\gamma(p,r,m)} ,
\label{eq:SFT} 
\end{equation}
where $p$ is the probability that one is adding $m$ new edges to the
system, $r$ is probability that one is rewiring $m$ edges and
$1-p-r$ is probabibility that one is adding a new node to the system.
The $\kappa$ and $\gamma$ in eq.(\ref{eq:SFT}) are given by \cite{AB}:
\begin{equation}
\kappa(p,r,m) = A(p,r,m) + 1 \qquad {\rm and}
                             \qquad \gamma(p,r,m) = B(p,r,m) + 1 ,
\end{equation}
where, in turn,
\begin{eqnarray}
A(p,r,m) &=& (p-r)\left[\frac{2m(1-r)}{1-p-r} + 1\right], \nonumber\\
B(p,r,m) &=& \frac{2m(1-r) + 1 - p - r}{m} .\label{eq:AB}
\end{eqnarray}
One can now write formally eq.(\ref{eq:SFT}) in the form resembling
eq.(\ref{eq:Pq}), i.e., as 
\begin{equation}
P(k) \sim \left[ \kappa(p,r,m)\right]^{-\gamma(p,r,m)}\, \cdot
\left[\, 1\, -\, (1-q)\cdot \frac{k}{(1-q)\kappa(p,r,m)}
\right]^{\frac{q}{1-q}}  
\end{equation}
and identify $\gamma(p,r,m) = q/(q-1)$ and $\lambda_0 = (1-q)
\kappa(p,r,m)$, or
\begin{equation}
q = 1 + \frac{1}{B} \qquad {\rm and}
             \qquad \lambda_0 = \frac{A + 1}{B} . \label{eq:ABT}
\end{equation}
However, it must be noticed that, at least in the example considered
here, the limit $q=1$ cannot be achieved because the quantity $B$
above is finite (for all reasonable values of parameters \cite{AB}).
It means then that formula (\ref{eq:Pq}) is more general and captures
(by means of parameters $q$ and $\lambda_0$) some additional feature of
complex networks, not present in its simple formulation (as, for
example, given by eq.(\ref{eq:SFT})).\\

To summarize:  We have demonstrated that, in order to apply the
information theory approach to analysis of stochastic networks one
has to use the nonextensive Tsallis information entropy $S_q$
\cite{T} leading to distribution $P_q(k)$ as given by
eq.(\ref{eq:Pq}). As shown in Fig. 1, such distribution provides 
satisfactory description of data on number of links in random
networks {\it in the whole range of variable $k$} by means of only
two parameters: the mean value of $k$ and the parameter
characterizing the type of information entropy to be chosen, $q$.
In this way one describes such disparate situations as the
exponential model ER \cite{ER} (for $q=1$) and the scale-free,
power-like models \cite{AB,BAJ} (with $q=\gamma/(\gamma-1)$). For the
value of nonextensivity parametr $q=3/2$, for which variance of our
system is divergent, one obtains the exponent $\gamma = 3$, which
seems to be limited value observed in analyses of diverse systems
displaying complex topology. Although only one example has 
been shown here in Fig. 1, it is obvious that one can just as easily 
also fit other, similar results discussed in the literature (cf., for
example, \cite{AB,DM}) \footnote{For the most recent application of
Tsallis statistics to investigation of Internet traffic problems see
\cite{Abe}, for the previous one see \cite{TA}.}. The other point is
the possible systematics of the $q$ parameter emerging from such a
search, but this problem is outside of the scope of our
presentation.\\ 

We conclude by saying that from the information theory point of view
eq. (\ref{eq:Pq}) could be used to fit different data providing a
pair of numbers $(q,\lambda_0)$ for each example. All competing models
could be then checked for their ability to correctly reproduce these
$(q,\lambda_0)$ and all models reproducing them correctly should be
regarded as {\it equally good} from the point of view of distribution
$P(k)$ because, according to the philosophy of infromation theory
approach, they apparently contain the same amount of information
existing in data which have been used. To distinguish between them
further one would have to use some additional information contained
in other network measures like, for example, clustering coefficient,
distance between nodes, cycles or graph spectra.\\

\end{document}